\documentclass[10pt,amssymb,superscriptaddress,showkeys,twocolumn,prl]{revtex4-2}
\usepackage{graphicx}
\usepackage{dcolumn}
\usepackage{bm}
\usepackage{changes}
\usepackage{amsmath,amssymb}

\makeatletter
\def\widetext@rule{}

\begin{document}

\title{Hybrid Disclination Skin-topological Effects in Non-Hermitian Circuits}
\author{Boyuan Li$^{1}$}
\thanks{These authors contributed equally.}

\author{Zekun Huang$^{1}$}
\thanks{These authors contributed equally.}

\author{Wenao Wang$^{1}$}
\author{Jiaxi Wang$^{1}$}
\author{Yu~Chen$^{1}$}

\author{Shaojie Ma$^{3}$}

\author{Ce Shang$^{2}$}
\email{shangce@aircas.ac.cn}

\author{Tie Jun Cui$^{1}$}
\email{tjcui@seu.edu.cn}

\author{Shuo Liu$^{1}$}
\email{liushuo.china@seu.edu.cn}


\begin{abstract}
\begin{center}
\textit{
$^{1}$State Key Laboratory of Millimeter Waves, Southeast University, Nanjing 210096, China.\\
$^{2}$Aerospace Information Research Institute, Chinese Academy of Sciences, Beijing 100094, China.\\
$^{3}$Shanghai Engineering Research Centre of Ultra Precision Optical Manufacturing,
Department of Optical Science and Engineering, School of Information
Science and Technology, Fudan University, Shanghai 200433, China.
}
\end{center}
The bulk–disclination correspondence (BDC) is a fundamental concept in Hermitian systems that has been widely applied to predict disclination states. 
Recently, disclination states have also been observed and experimentally verified in non-Hermitian systems with $C_{6}$ lattice symmetry, 
where gain and loss are introduced to induce non-Hermiticity. In this Letter, we propose a non-Hermitian two-dimensional (2D)  Su–Schrieffer–Heeger (SSH) disclination model with skin-topological (ST) disclination states, and calculate its biorthogonal Zak phase. Together with the real-space disclination index, we predict the emergence of disclination states in a $C_{4}$-symmetric non-Hermitian lattice and the corresponding fractional charge. We also generalize the symmetry indicator within the biorthogonal framework to predict the anomalous filling near the disclination core. Experimentally, the model is implemented on a nonreciprocal circuit platform, where we analyze the impedance matrix characterized by complex eigenfrequencies and directly observe the ST disclination states. Our work further extends the bulk–disclination correspondence to the non-Hermitian realm.
\end{abstract}
                              \maketitle
\emph{Introduction.---} 
In classical conditions of topological phases, hermiticity guarantees a real energy spectrum and ensures the robustness of the defined topological invariants. However, in photonic systems \cite{PhysRevLett.100.103904,ElGanainy2018NatPhys,Regensburger2012Nature,Jing2014PRL,Hodaei2014Science,Peng2014NatPhys,Wu2019Science,ElGanainy2018NatPhys2} and open quantum systems \cite{Minganti2019PRA}, the ubiquitous presence of gain and loss makes these conditions difficult to satisfy, thus necessitating the study of non-Hermitian topological phases \cite{PhysRevLett.121.136802,PhysRevLett.122.076801,PhysRevLett.124.086801,PhysRevLett.125.126402,PhysRevLett.124.056802,PhysRevB.102.205118,PhysRevB.102.241202,Yuce2019PRA}. 
Compared with Hermitian systems, non-Hermitian lattices support richer spectra and unconventional boundary responses, including the non-Hermitian skin effect (NHSE) \cite{Yao2018PRL,Kunst2018PRL,Yokomizo2019PRL,Song2019PRL,Okuma2020PRL,Borgnia2020PRL,Helbig2020NatPhys,Xiao2020NatPhys,Weidemann2020Science,Hofmann2020PRR} and hybrid skin-topological (ST) \cite{Lee2019PRL} boundary modes in which nonreciprocity and topology jointly dictate boundary localization. These features require the non-Bloch band description, formulated in the generalized Brillouin zone (GBZ) \cite{Yao2018PRL,Kunst2018PRL,Yokomizo2019PRL}.

Conventional polarization theory treats polarization as a local dipole density; however, it becomes inadequate for crystalline insulators where Bloch states are extended, and periodic boundaries make the position operator ill-defined. In contrast, modern polarization theory \cite{PhysRevB.48.4442,PhysRevLett.74.4035,PhysRevB.47.1651} adopts a global, geometric viewpoint, defining macroscopic polarization by the integral of the Berry connection—equivalently, the $U(1)$ Berry phase—accumulated over the occupied Bloch states across the entire Brillouin zone (BZ). Symmetry can quantize this geometric phase, so that a nontrivial polarization constitutes a bulk topological invariant. Equivalently, the nontrivial polarization indicates a quantized shift of the Wannier centers relative to the ionic background. Under bulk–boundary correspondence (BBC) \cite{RevModPhys.82.3045, RevModPhys.82.1959, RevModPhys.83.1057, RevModPhys.88.021004}, this mismatch gives rise to anomalous boundary charges, which can be equivalently characterized as a filling anomaly \cite{PhysRevB.99.245151, Li2020PRB,Peterson2020Science,Geier2018PRB,Xu2024PRB,Wada2025PRB}.

Geometric defects can couple to bulk topology and generate localized responses. Dislocations \cite{Ran2009NatPhys,PhysRevB.97.201111,Li2018NatComm,Nayak2019SciAdv,PhysRevResearch.3.033107,PhysRevB.104.L161106,PhysRevB.104.L241402} and disclinations \cite{Peterson2021Nature,PhysRevLett.124.243602,Liu2021Nature,PhysRevLett.127.066401} break local translational symmetry and thus induce localized electronic states and topological effects of abnormal quantum transport. 
Moreover, the bulk–disclination correspondence (BDC) establishes a direct link between bulk topology and disclination states and fractional charge near the disclination core.\cite{Ran2009NatPhys,Teo2013PRL,Ruegg2013PRL,Benalcazar2014PRB,Li2018NatCommun,Teo2010PRB,Li2020PRB,Liu2021Nature,Peterson2021Nature,Wang2020PRL,Wang2021NatCommun,Xue2021PRL,Geier2021SciPost,Deng2022PRL,Xie2022PRA,Lin2023NatRevPhys,Hwang2023NatPhoton, Peterson2021Nature,He2023FrontPhys}. In non-Hermitian settings, gain–loss has been demonstrated to induce topological disclination states \cite{Banerjee2024PRL}, while a complementary field-theoretic treatment connects disclinations to the NHSE \cite{PhysRevLett.127.066401}. However, there is a lack of a theoretical framework for predicting disclination states in non-Hermitian nonreciprocal systems.

In this paper, we theoretically establish and experimentally verify the non-Hermitian bulk--disclination correspondence in a non-Hermitian two-dimensional (2D) Su–Schrieffer–Heeger (SSH) disclination model. 
We find a topological invariant that incorporates both real-space and momentum-space information under non-Hermitian biorthogonal frameworks, which predicts the emergence of ST disclination states. We further explain these results by generalizing symmetry indicators to non-Hermitian biorthogonal settings.
Finally, we verify the above theory on an electrical circuit platform \cite{Imhof2018NatPhys,Lee2018CommPhys,Helbig2020NatPhys,Liu2020,Wang2020NatCommun,Liu2020PRApp,Liu2021Research,Shang2022AdvSci,Qiu2025NSR} and propose an experimental method for measuring complex eigenfrequencies that provide a direct mapping to the eigenstates of a non-Hermitian Hamiltonian.

\emph{Models.---} 
We start from the non-Hermitian 2D 
SSH model, in which each unit cell consists of four sites with nonreciprocal couplings along both the $x$ and $y$ directions, and reciprocal intercell couplings of strength $w$. In particular, we introduce unidirectional intracell couplings forming clockwise and counterclockwise loops denoted by  $v_p$ and $v_n$, respectively.
\begin{figure}[t!]
\includegraphics[width=\columnwidth]{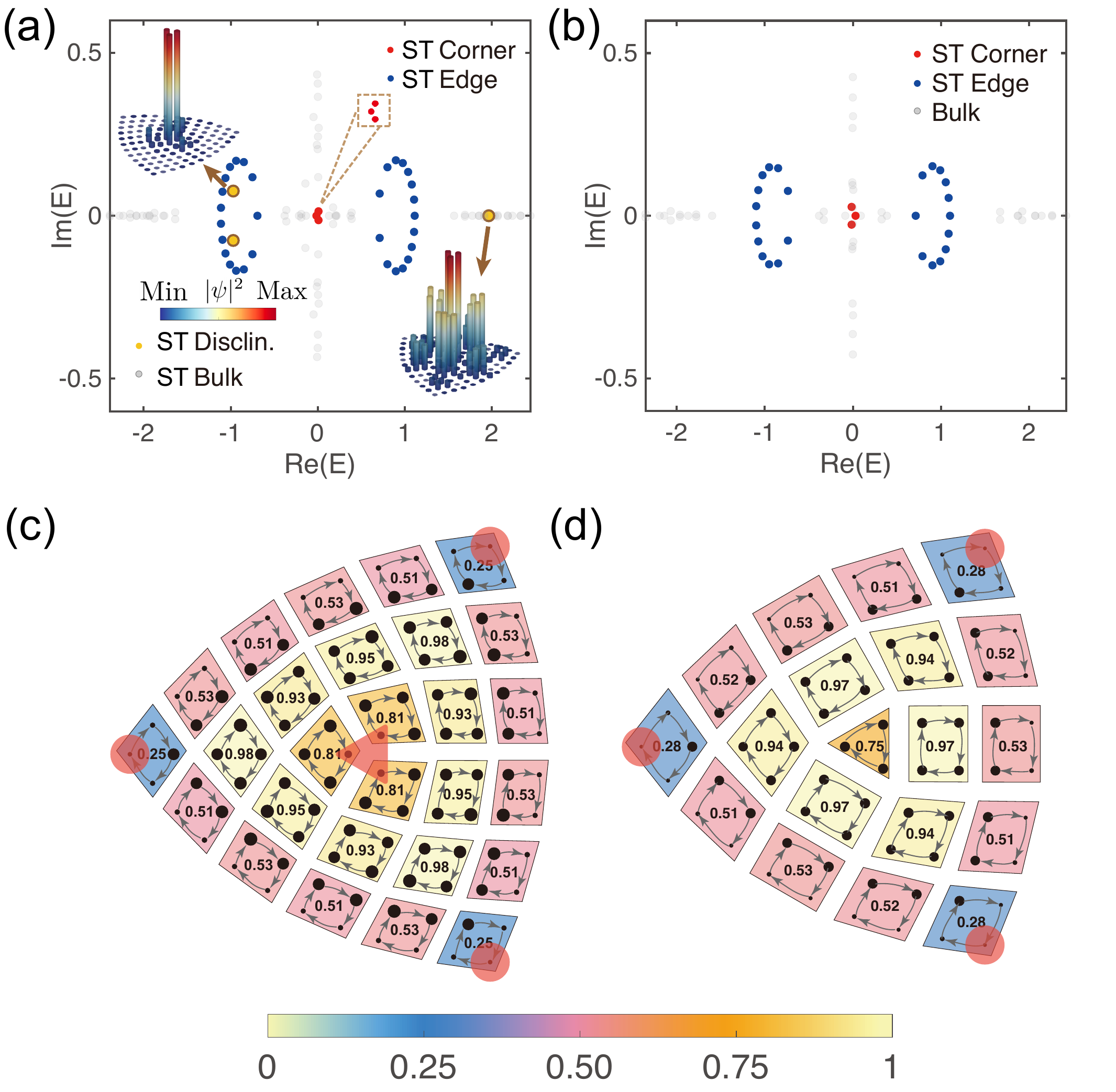}
\caption{
(a) Complex-energy spectrum for lattice size $l=12$ and the mode distribution, with corner modes (red), edge modes (blue), disclination modes (orange), and bulk modes (gray).
(b) Complex-energy spectrum for $l=10$.
(c-d) Density of states (DOS) of the occupied bands computed in the biorthogonal framework for (c) $\bm{s}=(0,0)$, (d) $\bm{s}=(1/2,1/2)$.
Red markers indicate the presence of bound states at the corresponding site positions. Parameters for all cases are $v_p=0.5, v_n=0, w=1$.}
\label{fig1}
\end{figure}
Next, we introduce a disclination with a frank angle $\Omega = -\pi/2$ into the system (see the End Matter for details). There exists a pair of complex conjugate
ST disclination modes (orange) together with a single purely
real mode (red) for even $\bm{|B|}$, [Fig. 1(b)], while these disclination states are absent when $\bm{|B|}$ is odd [Fig.~\ref{fig1}(c)]. Both types of disclination trap quantized fractional charge near disclination cores [Fig.~\ref{fig1}(d-e)], despite the presence of nonreciprocity. These ST disclination states and their fractional charge are intimately linked to the bulk topological characteristics of the corresponding model without disclinations. We first study the case without disclination states.

\begin{figure}[htp!]
\includegraphics[width=\columnwidth]{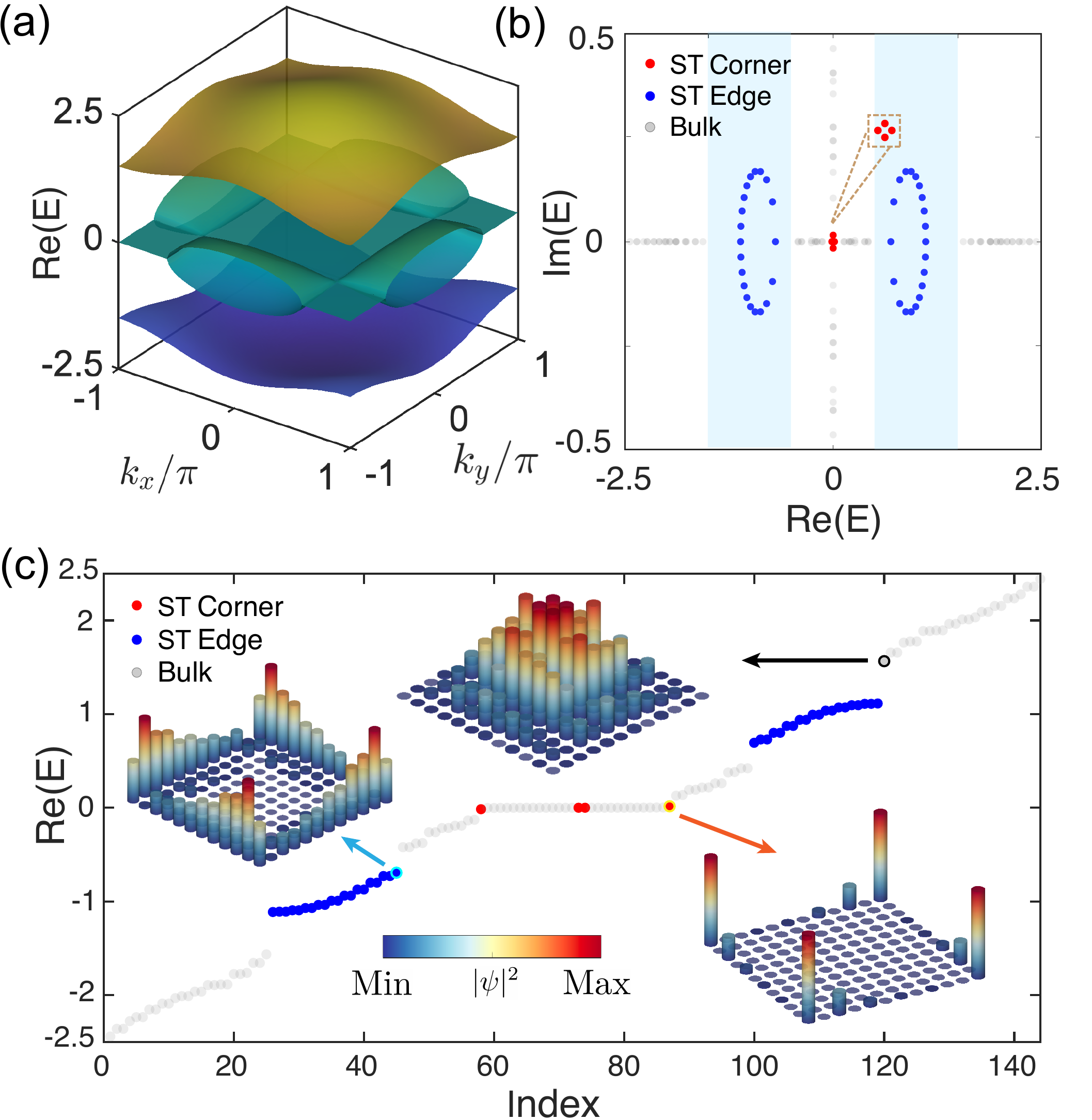}
\caption{Properties of the 2D non-Hermitian SSH model without disclinations when $v_p=0.5, v_n=0, w=1$.
(a) The real band structures under x-PBC, y-PBC.
(b) Complex spectrum of the $12\times12$ system under double OBC conditions.
Four ST corner modes appear at the center, and twenty ST edge modes are located in the line gap (blue shading).
(c) Spectrum of real part and the corresponding mode distributions.
}
\label{fig2}
\end{figure}

Figure~\ref{fig2}(a) shows the band structure under PBC along both directions, where there is a finite real band gap. The model exhibits only line gaps \cite{SM} and therefore does not host any nontrivial point gap topology  \cite{SM}. The complex energy spectra under $x$-open boundary condition (OBC) and $y$-OBC, and the corresponding mode distributions are shown in Fig.~\ref{fig2}(b-c).
We identify four ST corner modes, as well as twenty ST edge modes residing in the non-Hermitian line gap, whose number is determined by the finite lattice size.
Their emergence is jointly governed by the intrinsic topology and the boundary localization direction set by NHSE \cite{Li2025PRB, Lee2019PRL}.

\begin{figure}[b!]
\includegraphics[width=0.97\columnwidth]{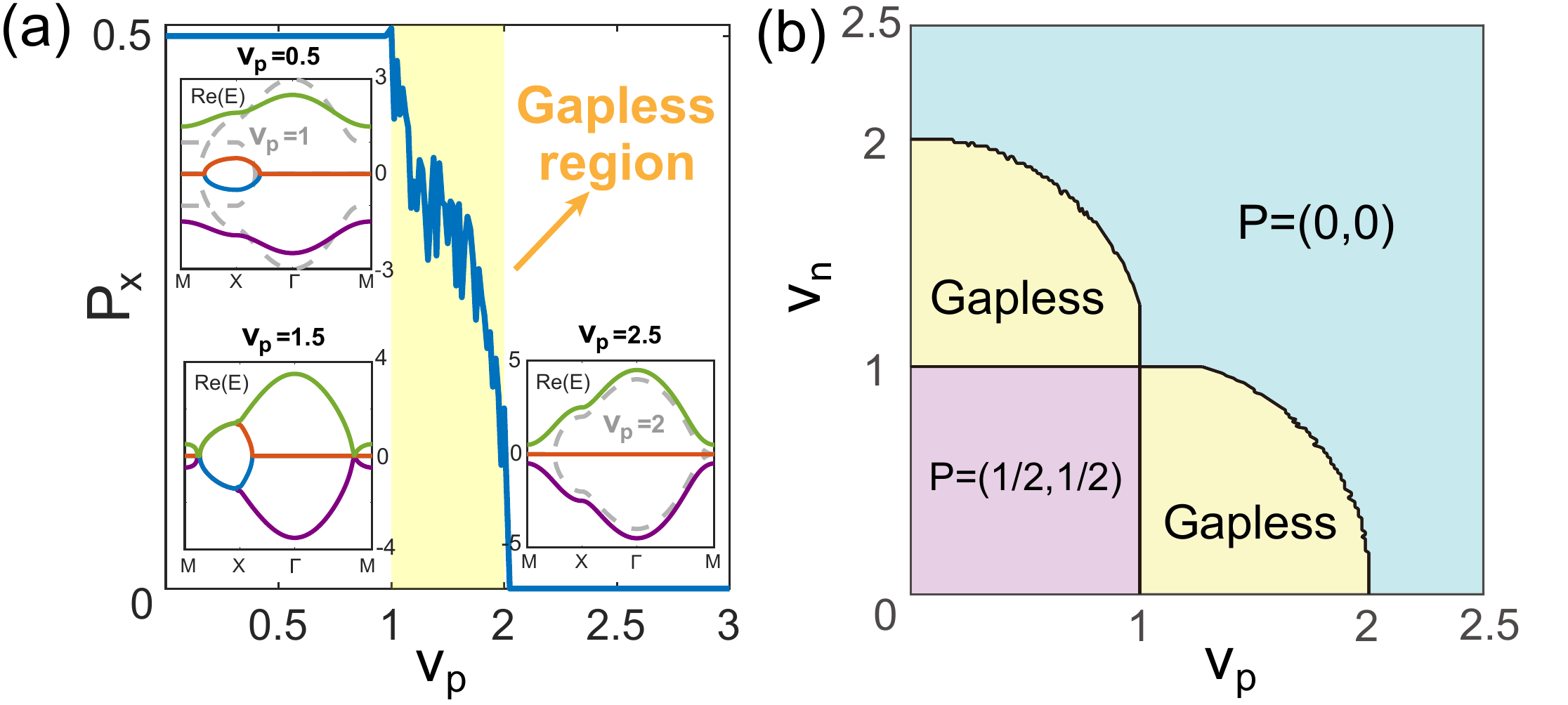}
\caption{(a) Polarization along the $x$-direction, $P_x$ ($v_n=0$). 
The system becomes gapless in the interval $1 \le v_p \le 2$. 
The band structures are shown for 
$v_p = 0.5$ (upper left, with the gray dashed curve indicating $v_p=1$), 
$v_p = 1.5$ (lower left), and 
$v_p = 2.5$ (right, with the gray dashed curve indicating $v_p = 2$).
(b) Phase diagram in the  $(v_p,v_n)$ plane when $w=1$.
}
\label{fig3}
\end{figure}

In this regime, the line gap structure allows the Hamiltonian to be continuously
deformed into an effectively Hermitian one without closing the gap. As a result, we can define a biorthogonal polarization vector
$\bm{\mathcal{P}}=(\mathcal{P}_x,\mathcal{P}_y)$ as a topological invariant associated with the geometric
phase. In the presence of crystalline symmetries, the biorthogonal polarization is quantized.
Mirror symmetries $M_{x,y}$ quantize $\mathcal{P}_i$ ($i=x,y$) to
$\mathcal{P}_i \equiv -\mathcal{P}_i\ (\mathrm{mod}\ e)$, leading to a $\mathbb{Z}_2 \oplus \mathbb{Z}_2$
classification. If only $C_4$ symmetry is preserved, as in our model, it enforces $\mathcal{P}_x=\mathcal{P}_y$ and reduces the invariant to a single $\mathbb{Z}_2$.
The biorthogonal Berry connection for the occupied bands is defined in terms of the left and right eigenstates as \(i\langle u_n^{L}|\partial_{k_i}u_n^{R}\rangle\). Accordingly, the biorthogonal polarization $\mathcal{P}_x$ can be formulated as
\begin{equation}
\mathcal{P}_x
= \frac{e}{(2\pi)^2}\int_{-\pi}^{\pi}dk_y\,
\arg\!
\prod_{n=1}^{N}
\frac{\langle u^{L}(k_{x,n},k_y)|u^{R}(k_{x,n-1},k_y)\rangle}
{\sqrt{\langle u^{L}_{n}|u^{R}_{n}\rangle\,
       \langle u^{L}_{n-1}|u^{R}_{n-1}\rangle}}.
\label{eq:px}
\end{equation}
Here, $N$ denotes the number of discrete sampling points used to
partition the Brillouin zone along the $k_x$ directions, respectively. Because the system respects $C_{4}$ symmetry, we have $\mathcal{P}_x=\mathcal{P}_y$. 
The results of 
biorthogonal polarization $\mathcal{P}_x$ are shown in Fig.~\ref{fig3}(a). 
In the regime $0<v_p<1$, the polarization is $\bm{\mathcal{P}}=(1/2, 1/2)$, 
whereas for $v_p>2$, it becomes $\bm{\mathcal{P}}=(0,0)$. 
As shown by the upper-left inset of Fig.~\ref{fig3}(a), when $v_p=0.5$, a finite band gap opens at the high-symmetry points (HSPs). 
When $v_p=1$ (gray dashed line), the two upper bands intersect, and the global real and imaginary \cite{SM} line gap no longer exists. The system remains gapless for $1<v_n<2$ (lower-right inset of Fig.~\ref{fig3}(a)).
As $v_{p}$ further increases, the system reaches the critical point for reopening the energy gap at $v_p=2$ (lower-right inset of Fig.~\ref{fig3}(a)).
At $v_p=2.5$, the gap is fully opened.
We further consider the more general case with $v_n\neq 0$, and obtain the phase diagram shown in Fig.~\ref{fig3}(b). When $0<v_n<1$ and $0<v_p<1$ (purple region), $\bm{\mathcal{P}}=(1/2, 1/2)$. The yellow region corresponds to the gapless phase in which the polarization vector no longer exists. In the blue region, the energy gap reopens, $\bm{\mathcal{P}}=(0, 0)$.

Next, we investigate the case after introducing a disclination.
According to homotopy theory, 
the equivalence class of the Burgers vector $\bm{B}$ forms a $\mathbb{Z}_{2}$ group when $\Omega = \pm \pi/2$, 
and a $\mathbb{Z}_{2} \oplus \mathbb{Z}_{2}$ group when $\Omega = \pm \pi$.
Since the parity of $2\bm{B}$ carries a geometric meaning, the disclination index $\boldsymbol{s}$ is introduced as a real-space topological invariant, defined by 
$2\boldsymbol{s} = (2\bm{B}) \bmod 2$ \cite{He2023FrontPhys}. 
The conventional bulk–disclination correspondence theory applies only to Hermitian systems. Here, combined with the biorthogonal polarization vector $\bm{\mathcal{P}}$, we extend this correspondence to non-Hermitian systems and define a new topological invariant $\boldsymbol{K}=(\bm{s} + \bm{\mathcal{P}})\bmod 1$, which can indicate the presence of ST disclination states.

In our model with $C_{4}$ rotational symmetry, the relation between the lattice size $l$ and $\boldsymbol{s}$ is given by 
$s_{x}=s_{y}=\tfrac{1}{4}(l \bmod 4)$. 
Hence, for finite lattices whose $l \bmod 4 = 0$, $\boldsymbol{s} = (0, 0)$, 
and for $l \bmod 4 = 2$, $\boldsymbol{s} = (1/2, 1/2)$.
Fig.~\ref{fig1}(b) shows the complex energy spectrum of the disclination lattice with $l=12$ , corresponding to $\boldsymbol{K}=(1/2,1/2)$. 
A pair of conjugate disclination states emerges within the loop of the ST edge state on the negative real axis, and an additional pure disclination state is embedded with the bulk states on the positive real axis.
In contrast, fig.~\ref{fig1}(c) shows the case of $l=10$, corresponding to $\boldsymbol{K}=(0,0)$, where no ST disclination state is observed.

The fractional charge is also essential, as it is equivalent to the anomalous filling of the Wannier center.
From the biorthogonal polarization vector, we infer that the Wannier center resides at the Wyckoff position \textbf{1a}, implying anomalous filling at the corners and at the disclination core. 
We confirm this by the biorthogonal density of states (DOS) within the occupied bands of the disclination
model, 
\begin{equation}
Q_{d}=\sum_{m}\sum_{n}\Big|\langle\psi^{L}_{n} | m\rangle\langle m|\psi^{R}_{n}\rangle\Big|\ (\bmod 1).
\end{equation}
Here, the summation over $m$ runs over the sites surrounding the disclination core, while $n$ traverses all occupied bands. 
The biorthogonal density of states (DOS) is shown in Fig.~\ref{fig1}(d-e). For $\boldsymbol{s}=(0,0)$, the result is $Q_{d}=0.75$.
For the case with $\boldsymbol{s}=(1/2,1/2)$, the fractional charge per unit cell near the disclination core, which is $Q_{u}=0.81$, slightly deviates from the theoretical value due to the limited lattice size \cite{SM}. In the thermodynamic limit, the total fractional charge near the disclination core is $Q_{d}=(3Q_u)\bmod1\approx0.25$.

\begin{figure*}[ht]
\includegraphics[width=0.95\textwidth]{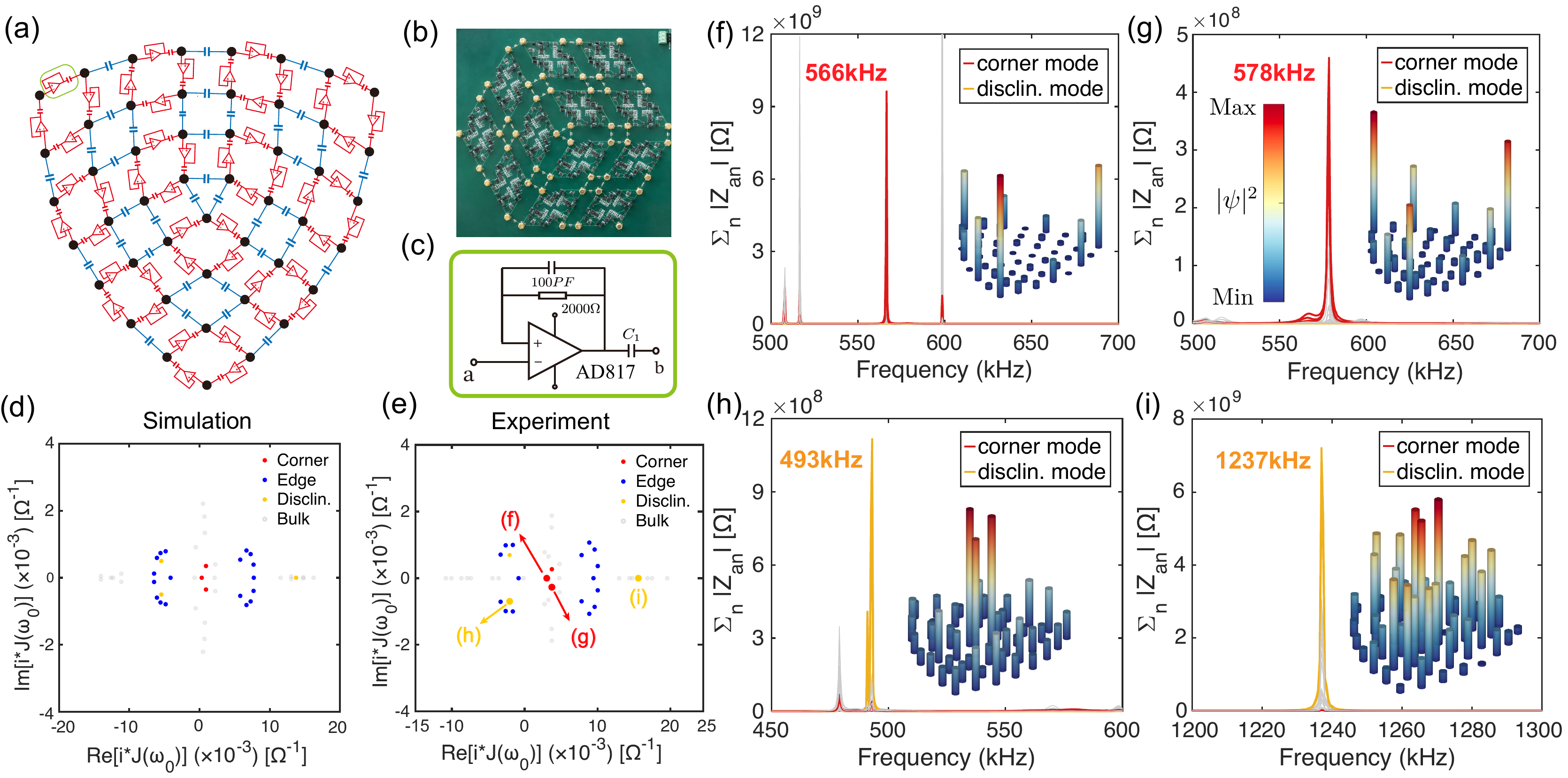}
\caption{Experimental results of an $8 \times 8$ disclination circuit with Frank angle $-\pi/2$. (a) Schematic of the nonreciprocal disclination circuit. Red components indicate voltage followers connected to a terminal capacitor $C_{1}$, and blue components denote coupling capacitors $C_{2}$. 
(b) The picture of the disclination circuit board. 
(c) Internal structure of the voltage follower.  
(d) Simulated, (e) experimental results of the spectrum of $i\bm{J}$ at the resonance frequency $531\text{kHz}$. 
(f) Case of $\sigma_g = 0$. A resonance peak appears at $566\text{kHz}$, corresponding to a ST corner state located on the real energy axis. The right inset shows the distribution of the right eigenstate amplitude. 
(g) Case of $\sigma_g = 0.0002697S$. A conjugate ST corner mode with negative imaginary part appears at $578\text{kHz}$. 
(h) Case of $\sigma_g = 0.0006255S$. A conjugate ST disclination mode with negative imaginary part appears at $493\text{kHz}$.
(i) Case of $\sigma_g = 0$. An ST disclination state with purely real eigenfrequency is observed at 1237kHz.}
\label{fig4}
\end{figure*}

\emph{Biorthogonal symmetry indicators.---} 
Here, we generalize the Hermitian rotation invariants to a biorthogonal non-Hermitian platform to explain the results.
For a 2D insulator with $C_n$ rotation symmetry \cite{Li2020PRB}, the Bloch Hamiltonian satisfies
$\hat r_n H(\bm{k})\hat r_n^\dagger=H(R_n\bm{k})$, where $\hat r_n$ is the rotation operator corresponding to a rotation by $2\pi/n$.
In a Hermitian system, at rotation-invariant HSP $\Pi$ with $R_n\Pi=\Pi$, one has $[\hat r_n,H(\Pi)]=0$.
Whereas in the non-Hermitian case, the occupied subspace is spanned by biorthogonal left and right Bloch eigenstates satisfying $\langle u^{l}_{L}(\Pi)|u^{m}_{R}(\Pi)\rangle=\delta_{lm}$.
Accordingly, the $C_n$-rotation eigenvalues at $\Pi$ are obtained by diagonalizing the biorthogonal rotation matrix
\begin{equation}
\mathcal{R}^{\Pi}_{lm}=
\bigl\langle u^{l}_{L}(\Pi^{(n)})\big|\hat r_n\big|u^{m}_{R}(\Pi^{(n)})\bigr\rangle,
\quad l,m\in \text{occ.},
\end{equation}
whose eigenvalues define $\Pi_{p,NH}^{(n)}$ ($p=1,\dots,n$).
This enables the definition of symmetry indicators purely from band representations,
\begin{equation}
\bigl[\Pi_{p,NH}^{(n)}\bigr]\equiv \#\Pi_{p,NH}^{(n)}-\#\Gamma_{p,NH}^{(n)},
\end{equation}
where $\#\Pi_{p,NH}^{(n)}$ counts the number of occupied bands at HSP $\Pi$ with the eigenvalue $\Pi_{p,NH}^{(n)}$, and $\Gamma$ is the center of the BZ.
Symmetry-indicator relates rotation invariants in momentum space to the generalized Wannier-orbital content in real space, so that the fractional charge bound to a disclination can be written as an index depending solely on $\{[\Pi_{p,NH}^{(n)}]\}$.
For $C_4$ symmetry, the fractional charge bound to a disclination is given by \cite{Li2020PRB}
\begin{equation}
 Q_{d} = \frac{\Omega}{2\pi}\Bigl( [{X}_{1, NH}^{(2)}] + \frac{3}{2}[{M}_{3,NH}^{(4)}] - \frac{1}{2}[{M}_{1,NH}^{(4)}] \Bigr) + \bm{T}\cdot\bm{\mathcal{P}} \bmod 1,
\label{8}
\end{equation}
where $\Omega$ is the Frank angle, $[X_{1,NH}^{(2)}]$ and $[M_{1,3,NH}^{(4)}]$ are the biorthogonal rotation invariants. 
Here $\bm{T}\!\cdot\bm{\mathcal{P}}$ captures the translational part of the holonomy, with $\bm{T}$ the normal direction associated with this translation \cite{Li2020PRB}. In our model, we obtain $[X_{1,NH}^{(2)}]=[M_{3,NH}^{(4)}]=-1$, and $[M_{1,NH}^{(4)}]=1$.
Specifically, for $l \bmod 4 = 2$, $(\bm{T}\!\cdot\!\bm{\mathcal{P}}) \bmod 1 = 0$, leading to $Q_d  = 3/4$.
For $l \bmod 4 = 0$, $(\bm{T}\!\cdot\!\bm{\mathcal{P}}) \bmod 1 = 1/2$,  and hence $Q_d = 1/4$, which exactly coincide with the numerical results.

\emph{Experiments.---} 
Hamiltonian can be mapped to the circuit Laplacian. In experiments, we fabricate an 8×8 circuit lattice with a Frank angle of $-\pi/2$, shown in Fig.~\ref{fig4}(a-b).
 The ideal circuit Lagrangian matrix is given by
$\bm{J}(\omega) = i \omega \bm{C} - (i/\omega) \bm{W}$, where $\bm{C}$ and $\bm{W}$ are
the matrices of capacitance and inverse inductance. When driven at the resonant frequency, the circuit Laplacian $\bm{J}(\omega_{0}) = i \omega_{0} C_{t}\bm{H}$ is linearly related to the Hamiltonian of
the system, where $C_{t}=C_{1}/v_{p}$. The circuit parameters are chosen as $C_{1}=1\text{nF}$, $C_{2}=2\text{nF}$ and $L_{g}=18\mu\text{H}$, in both simulations and experiments, leading to the on-site resonant frequency $\omega_0=531$kHz.
The nonreciprocal coupling is realized via voltage followers [Fig.~\ref{fig4}(c)] employing operational amplifiers (AD817) capable of driving sufficiently large capacitive loads.
  
Next, we express the Hamiltonian in the biorthogonal basis as $
\bm{H} = \bm{V_R} \bm{\Lambda} \bm{V_L}^{\dagger}
$, where \( \bm{\Lambda} \) denotes the diagonal matrix of eigenvalues, and \( \bm{V_R} \) and \( \bm{V_L} \) collect the right and left eigenvectors, respectively. The impedance matrix then takes the following form:
\begin{equation}
Z_{ab}
= \sum_{n} \frac{1}{i\omega C_t}\,({V_R})_{an}
\frac{1}{\lambda_{n} - \lambda(\omega)}
({V_L}^{\dagger})_{nb},
\end{equation}
where $\lambda(\omega) = -C_g/C_t + 1/(\omega^{2} C_t L_g)$ and $\lambda_n$ is the $n$-th eigenvalue of the Hamiltonian.
When the circuit is driven at an eigenfrequency $\omega_n$ corresponding to the n-th mode of Hamiltonian, the eigenstates of the electrical circuit can be mapped exactly onto those of Hamiltonian via the following relations: $|Z_{a1}(\omega_n)| \propto |\psi_{R}^{na}|, |Z_{1a}(\omega_n)| \propto |\psi_{L}^{na}|, |Z_{aa}(\omega_n)| \propto |\psi_{R}^{na}\psi_{L}^{na}|$. Here, we choose the sum of the magnitudes of each row of the impedance matrix to characterize the corresponding eigenstate $|\psi_{R}^{na}|$.

However, in non-Hermitian circuits, it should be noted that the eigenfrequencies are generally complex. Instead of exciting the circuit at a complex frequency, we introduce a compensating conductance term $\sigma_g(\omega_n)$ grounded at each circuit node (see the End Matter for details),
\begin{equation}
 \sigma_g(\omega_n) = -\frac{2\omega_{nr}\omega_{ni}}{(\omega_{nr}^2+\omega_{ni}^2)\sqrt{\omega_{nr}^2-\omega_{ni}^2}L_g}.
 \label{10}
\end{equation}
$\omega_{nr}$ and $\omega_{ni}$ are real and imaginary parts of $\omega_n$.
By exciting the circuit at a real frequency $\omega_{\mathrm{eff}}=|\omega_n^2|/\sqrt{Re(\omega_n^2)}$, we can effectively access the eigenstate of the n-th complex eigenmode $\omega_n$.
The detailed derivation and verification by ADS simulations are provided in \cite{SM}.

Fig.~\ref{fig4}(d-e) shows, respectively, the simulated and experimental spectrum of the circuit Laplacian at the resonant frequency 531kHz. By introducing the appropriate compensating conductance term $\sigma_g(\omega_n)$ into the diagonal of the circuit Laplacian, we obtain the corresponding impedance spectra and the mode distributions at each resonance frequency $\omega_n$, as shown in Fig.~\ref{fig4}(f-i).
With $\sigma_g=0$, we observe two real modes at 566kHz and 1237kHz that correspond, respectively, to the ST corner mode [Fig.~\ref{fig4}(i)] and the ST disclination mode [Fig.~\ref{fig4}(f)].
To access the complex eigenmode with a negative imaginary part, an additional conductance $\sigma_g = 0.0006255S$ and $\sigma_g = 0.0002697S$ is manually added to the diagonal of $i\bm{J}(\omega_0)$. This procedure shifts the imaginary parts to zero, resulting in pronounced impedance resonance peaks at 578kHz and 493kHz, which correspond, respectively, to the ST corner mode [Fig.~\ref{fig4}(g)] and ST disclination mode [Fig.~\ref{fig4}(h)]. 

\emph{Conclusion.---} 
 We establish a non-Hermitian 2D SSH disclination model that simultaneously hosts ST corner, ST edge, and ST disclination modes, and extend the BDC to the non-Hermitian framework. In such cases, the reciprocal-space topological invariant should be replaced by the biorthogonal Zak phase. Meanwhile, we formulate the symmetry indicators within the non-Hermitian biorthogonal framework, enabling an exact prediction of the fractional charge near the disclination core. 
In the experiment, we establish a direct correspondence between the impedance matrix and the non-Hermitian biorthogonal eigenvector, and propose a method for resolving the complex eigenfrequencies in non-Hermitian electrical circuits. This allows us to experimentally observe the ST corner and ST disclination modes in the complex frequency plane.

\section*{Acknowledgements}
This work acknowledges funding from the National Key Research and Development Program of China under Grant No.\ 2022YFA1404903, the National Natural Science Foundation of China under Grant No.\ 62288101, and Project Nos. E4BA270100, E4Z127010F, E4Z6270100, E53327020D of the Chinese Academy of Sciences.

\begin{figure}[b!]
\includegraphics[width=\columnwidth]{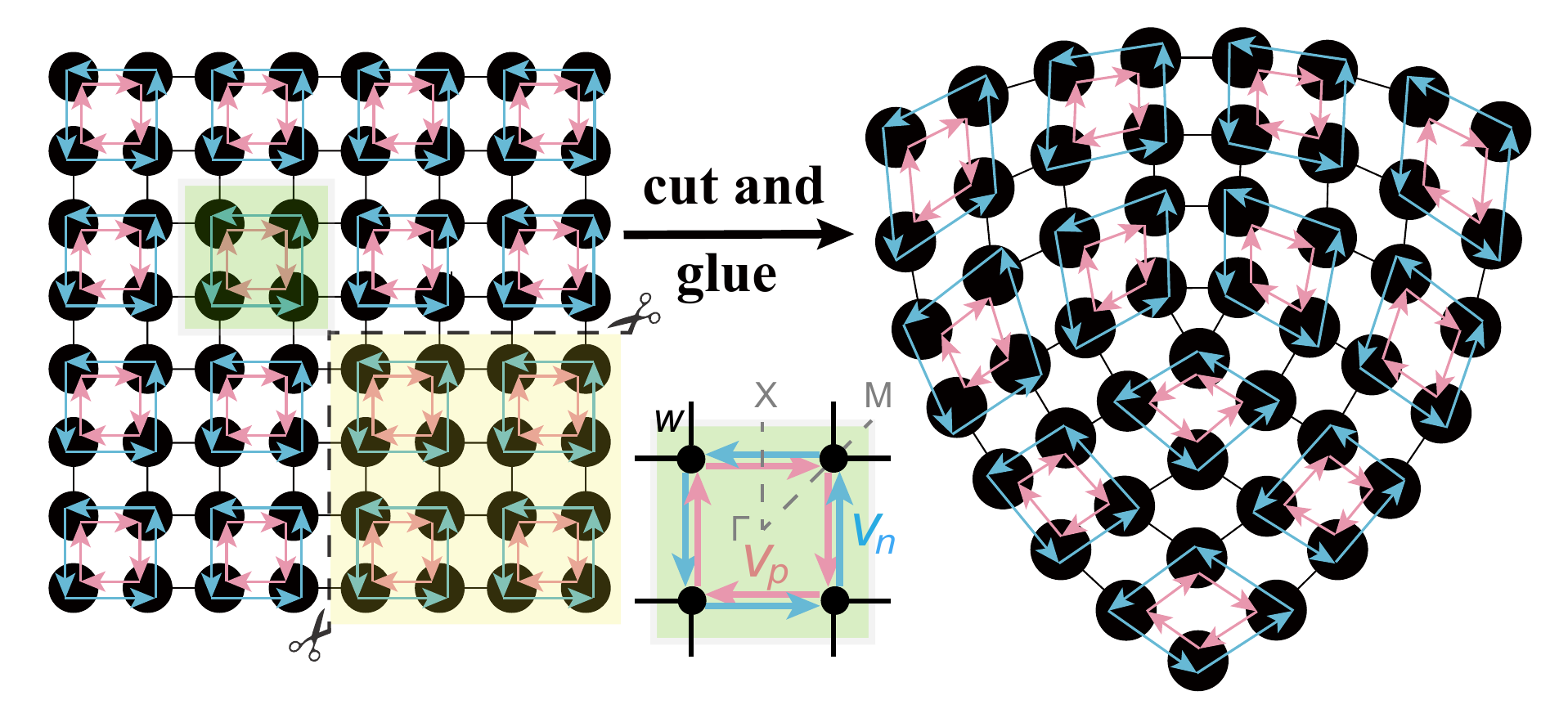}
\caption{(a) Schematic of an $8\times 8$ non-Hermitian 2D SSH lattice with a disclination of Frank angle $-\pi/2$. Clockwise intracell couplings (pink) are $v_p$ and counterclockwise couplings (blue) are $v_n$.
The lower-right quarter is cut out, and the exposed edges are glued by the intercell coupling $w$. 
}
\label{fig5}
\end{figure}

\bibliography{Disclination}
\clearpage
\onecolumngrid
\begin{center}
\textbf{\large End Matter}
\end{center}
\twocolumngrid
\emph{The non-Hermitian 2D SSH model---} 
We can write the Hamiltonian as follows:
\begin{equation}
\begin{split}
H(k_x,k_y)
&= \bm{\sigma}_0 \otimes \big[(v+w\cos k_x)\bm{\tau}_1 + (w\sin k_x)\bm{\tau}_2\big]\\
&+ \bm{\sigma}_3 \otimes \big[(-i\delta)\bm{\tau}_2\big]
+ (v+w\cos k_y)\big(\bm{\sigma}_1 \otimes \bm{\tau}_0\big)\\
&+ \bm{\sigma}_2 \otimes \big[(w\sin k_y)\bm{\tau}_0 + i\delta\bm{\tau}_3\big].
\end{split}
\end{equation}

where \(v= (v_p+v_n)/2 \) and \(\delta=(v_p-v_n)/2\), \(\boldsymbol{\sigma_i}\) and \(\boldsymbol{\tau_i}\) are Pauli matrices with \(i = 0, 1, 2, 3\). 
By cutting and gluing the lower-right quadrant of a square lattice with 4×4 unit cells, as illustrated in  Fig.~\ref{fig5}, we introduce a disclination with a frank angle $\Omega = -\pi/2$ into the system.

\emph{Effective Protocol for Probing Complex eigenmodes---} We establish the correspondence between non-Hermitian systems and nonreciprocal circuit parameters. We also present the treatment of imaginary
frequencies, deriving an effective real frequency $\omega_{\mathrm{eff}}$  and compensating conductance $\sigma_g$, mentioned in the main text. 
Firstly, we begin with the basic formalism of the circuit Laplacian,
\begin{equation}
\bm{J}(\omega) = i\omega \bm{C} + \frac{\bm{W}}{i\omega}
  = i\omega\bigl(-C_g \bm{I} + C_t \bm{H}\bigr) - \frac{\bm{I}}{i\omega L_g},
\end{equation}
which is equal to,
\begin{equation}
\bm{J}(\omega)=i\omega C_t(\bm{H}-\lambda(\omega)\bm{I}),\quad
\lambda(\omega) = \frac{C_g}{C_t} - \frac{1}{\omega^{2} L_g C_t}.
\label{S6}
\end{equation}

Since the Hamiltonian is non-Hermitian, we can expand it in terms of left and right eigenvectors as
 $
\bm{H} = \bm{V_R} \bm{\Lambda} \bm{V_L}^{\dagger}
$,
then we obtain:
\begin{equation}
\bm{J}(\omega)
= i\omega C_t \bm{{V_{R}}}\!\left(\bm{\Lambda}-\lambda(\omega)\bm{I}\right)\bm{{V_{L}}}^{\dagger}.
\end{equation}
By inverting the circuit Laplacian above, the impedance matrix is obtained as follows,
\begin{equation}
Z_{ab} 
=\sum_{n} \frac{1}{i\omega C_t}\,({V_R})_{an}
\frac{1}{\lambda_{n} - \lambda(\omega)}
({V_L}^{\dagger})_{nb}.
\label{eq:Zab}
\end{equation}
where $\Lambda_{mn}=\lambda_n\delta_{mn}$, and we also use the biorthogonal relation
$
 \bm{V_L}^{\dagger}\bm{V_R} =\bm{I}
$. 
Here $V_R=[|u_n^{R}\rangle]$ and $V_L=[|u_n^{L}\rangle]$ collect the left and right eigenvectors with
$H|u_n^{R}\rangle=\lambda_n|u_n^{R}\rangle$ and $\langle u_n^{L}|H=\langle u_n^{L}|\lambda_n$.
When $\lambda(\omega_n)=\lambda_n$, the denominator approaches zero, resulting in a pronounced impedance peak.
Through this formulation, 
the eigenstates of the electrical circuit can be mapped
exactly onto those of the Hamiltonian via the relations:
$
|Z_{a1}(\omega_n)| \propto |\psi_{R}^{na}|,
|Z_{1a}(\omega_n)| \propto |\psi_{L}^{na}|,
|Z_{aa}(\omega_n)| \propto |\psi_{R}^{na}\psi_{L}^{na}|.
$

Considering that the eigenfrequency generally takes complex values
$
\omega_n = \omega_{nr} + i\omega_{ni},
$
it is impossible to excite the system at purely real frequencies to get the resonance peak associated with the
corresponding mode. To overcome this limitation, we introduce a compensating conductance $\sigma_g(\omega_n)$. By substituting $\omega_n=\omega_{nr}+i\omega_{ni}$ into Eq.~(\ref{S6}), we can rewrite $\lambda(\omega_n)$ as:
\begin{equation}
\lambda(\omega_n)
=
\frac{C_g}{C_t}
-\frac{\omega_{nr}^2-\omega_{ni}^2}{(\omega_{nr}^2+\omega_{ni}^2)^2 L_g C_t}
+\frac{2 i\omega_{nr}\omega_{ni}}{(\omega_{nr}^2+\omega_{ni}^2)^2 L_g C_t}.
\label{eq:first_lambda}
\end{equation}

We then consider the circuit Laplacian with the compensating conductance  
$\boldsymbol{\sigma}=-\sigma_g\bm{I}$:
\begin{equation}
\bm{J}(\omega_n')=i\omega_n' C_t \bigl( \bm{H}-\lambda'(\omega_n')\bm{I} \bigr),
\label{eq:shifted_lambda}
\end{equation}
where $\lambda'(\omega_n')
=
{C_g}/{C_t}
-{1}/{(\omega_{n}'^{2} L_g C_t)}
-{i\sigma_g}/{(\omega_n' C_t)}.$
Comparing the real and imaginary parts of Eq.~\eqref{eq:first_lambda} with 
\eqref{eq:shifted_lambda}, we obtain:
\begin{equation}
\omega_{\mathrm{eff}}=\omega_{n}'=|\omega_n^2|/\sqrt{Re(\omega_n^2)},\qquad 
\label{solution}
\end{equation}
\begin{equation}
\sigma_g(\omega_n) = -\frac{2\omega_{nr}\omega_{ni}}{(\omega_{nr}^2+\omega_{ni}^2)\sqrt{\omega_{nr}^2-\omega_{ni}^2}L_g}.
\end{equation}
These two equations show that, after introducing compensating conductance, the required excitation frequency $\omega_n’$ is no longer simply given by the real part of the original eigenfrequency. Therefore, to observe the eigenmode associated with the $n$-th eigenfrequency, we could first compute the required compensating conductance theoretically, and
then use $
\bm{Z}^\prime(\omega_n) = (\bm{Z}^{-1}(\omega_n) + \sigma_g(\omega_n) )^{-1}
$ to process the measured data at the corresponding frequency.
Finally, we can observe the effective impedance peak corresponding to the target eigenstate. The corresponding ADS simulation results are provided in \cite{SM}.

\end{document}